\documentstyle[twoside,fleqn,espcrc2]{article}

\def\Jou#1#2#3#4{{#1} #2 (#4) #3 }

\def\NPB{{Nucl. Phys.} B}
\def\PLB{{Phys. Lett.}  B}

\def\PRL{Phys. Rev. Lett.}

\def\PRD{{Phys. Rev.} D}

\def\ZPC{{Z. Phys.} C}

\newfam\BMath
\font\BMathL=cmmib10 
\font\BMathl=cmmib7
\font\BMathm=cmmib5
\textfont\BMath=\BMathL \scriptfont\BMath=\BMathl
\scriptscriptfont\BMath=\BMathm

\def\mthbf#1{{\fam\BMath #1}}

\def\be{\begin{equation}}
\def\ee{\end{equation}}
\def\bea{\begin{eqnarray}}
\def\eea{\end{eqnarray}}
\def\eref#1{Eq. (\ref{#1})}

\def\a{\alpha}
\def\b{\beta}
\def\c{\chi}
\def\f{\phi}
\def\j{\psi}
\def\m{\mu}
\def\p{\pi}

\def\cm{{\cal M}}

\def\kt{{\mthbf k}_\perp}
\def\dd{\mbox{\rm d}}

\def\ot{\otimes}
\def\fx{\!\!\!\!}

\font\fiverm=cmr5

\newcommand{\AmS}{{\protect\the\textfont2
  A\kern-.1667em\lower.5ex\hbox{M}\kern-.125emS}}

\title{Colour octet contribution to exclusive P-wave charmonium
       decay into nucleon-antinucleon}

\author{S.M.H. Wong
        \thanks{ 
         This work is supported by the European Training and 
         Mobility of Researchers programme under 
         contract no. ERB-FMRX-CT96-0008}
        \address{\hskip -0.0cm
        Nuclear and Particle Physics Section, University of Athens, 
        Panepistimiopolis, GR-15771 Athens, Greece  \\
        Institute of Accelerating Systems and Applications (IASA),   
        P.O. Box 17214, GR-10024 Athens, Greece \\
        and \\
        Fachbereich Physik, Universit\"at Wuppertal, D-42097 Wuppertal, 
        Germany}
}

\begin{document}

\begin{abstract}
We show that although there is no infrared divergence in exclusive 
P-wave quarkonium decays, the colour octet contribution is no 
less important than in the inclusive decay. Results from more proper 
calculations with only colour singlet contribution are not sufficient 
to account for the measured partial decay widths and so the next 
higher Fock state must be included. Using the $\c_J$ decay into 
$N\bar N$ as an example, a scheme is devised to yield estimates of 
the decay widths using both contributions, the results are comparable 
with experimental measurements.
\hfill{\bf IASA 98-2, UA/NPPS-98-12}
\end{abstract}

\maketitle

\section{Introduction}

In the late 60's and early 70's, Barbieri et al \cite{bgk,bgr}
considered the inclusive decay of P-wave heavy quark-antiquark 
system into hadrons. Such decays proceed via the annihilation 
of the heavy fermions into massless gluons. For P-wave quarkonium, 
annihilation into two gluons is allowed and this is the leading
contribution to the inclusive hadronic decay width, at least for
the total angular momentum $J=0$ and $J=2$ system. For $J=1$ P-wave 
quarkonium, the decay into two massless spin-1 gluons is forbidden. 
The leading decay process in this case is therefore one gluon and
a light quark-antiquark pair. Whereas the kinematics of the 
leading decay process of the even-$J$ systems into two gluons 
is very simple and completely well defined, that of the $J=1$
system is not. In the rest frame of the quarkonium, the decay
into two gluons would proceed with them moving apart from 
each other with equal and opposite momentum. The three-body 
leading decay of the $J=1$ quarkonium is, on the other hand, 
kinematically less restrictive so the situation where the
quark and antiquark move apart with nearly equal and opposite
momentum with an accompanying very soft gluon is allowed. 
The corresponding probability on the tree graph level 
from such a process with a very soft gluon is easily shown to 
contain infrared divergence {\em when all the participants are
on their respective mass shells}. In view of the infrared divergence 
and the fact that the heavy quark and antiquark are really in
a bounded system, one should perhaps keep them off shell by
an amount corresponding to the binding energy $\varepsilon$
of the system. Thus using $\varepsilon$ as an infrared cutoff,
one gets for the $J=1$ quarkonium decay width
\be \Gamma_{J=1} \sim   \a^3_s \int_\varepsilon {{\dd q} \over q} 
                 \simeq \a^3_s \ln \varepsilon      \;\;\; .
\ee
This is the solution proposed in \cite{bgr}. In fact, this
infrared divergence also appears in the decay widths of the
even-$J$ P-wave quarkonia but at the next-to-leading order. 

With the more recent advances in quarkonium physics \cite{bbl1,bbl2},
we now know that this infrared divergence should not be present.
The physical picture of the quarkonium in the early 70's was
a heavy quark bounded with an equally heavy antiquark. 
The missing component in this picture is the higher Fock 
states of the quarkonium. The contribution to the decay 
into $gq\bar q$ from the so-called colour octet, the next 
higher Fock state of the P-wave quarkonium, becomes degenerate 
with that from the lowest valence state when the
gluon approaches the infrared. When both are included, there is 
no infrared divergence. This is the situation of the inclusive
P-wave quarkonium decay. Colour octet is needed and is introduced
to cancel infrared divergence. 

In exclusive decay, however, the heavy quark system decays 
from a bounded system into other 
bounded systems, the same infrared divergence in the inclusive 
decay does not appear. So there is no need for colour octet 
and the partial widths of the various exclusive decay modes 
have been calculated from the valence colour singlet component 
alone, see \cite{aa,coz} for example. We will show that, on 
the contrary, the colour octet is very important for
P-wave quarkonium.

\section{Angular momentum suppression of P-wave wavefunctions}

The decay of a quarkonium of mass $M$ through annihilation is 
a short distance process with the annihilation length given by
$L\sim 1/M$. For a heavy quarkonium, $1/M$ is very small so
we only need the quarkonium wavefunction near the origin
$\j (L\sim 0)$. This is true only for a S-wave quarkonium,
for a P-wave, the wavefunction at the origin is zero so
one expands the wavefunction around the orgin
and uses instead the quantity $L \j'(L\sim 0)$ for the
infrared confinement physics. In momentum space, this means 
\bea \mbox{S-wave: \hskip 1.0cm} \j_S (0) \mbox{\hskip 1.5cm}
     \fx & \longrightarrow & \fx \tilde  \j_S (k)             \\
     \mbox{P-wave: \hskip 1.0cm} \j_P (0) \sim L \j'(0)
     \fx & \longrightarrow & \fx \frac{k}{M} \tilde \j_P (k)  \; .
     \nonumber \\
\eea
Assuming that there is no significant difference between the 
S- and P-wave Fourier components $\tilde \j(k)$, which seems
reasonable, it follows that the P-wave wavefunction is weighed 
down by $1/M$ in comparison with that of the S-wave. Therefore 
P-wave is relatively suppressed on the level of the wavefunction. 
We will see that this is very important in the next section.

\section{Comparing colour singlet with octet}
\label{s:svso}

In order to find out whether the colour octet component in
exclusive quarkonium decay is negligible or not when compared 
with the colour singlet, we now perform power counting in the 
only large scale of the process, namely, $M$. We must choose
an explicit process and a scheme for the calculation of the 
decay probability amplitude in order to be able to do this.
We use $\c_J$ decay into $N\bar N$ as the example.
For the purpose of power counting, it is sufficient to choose 
the simpler standard hard scattering scheme of Brodsky and Lepage 
\cite{bl}. In this scheme, the probability amplitude is given by a 
convolution of distribution amplitudes $\f(x)$ and hard perturbative 
part $\hat T_H$.
\be \cm \sim f_{\c_J} \f_{\c_J}(x) \ot f_N \f_N(x) 
    \ot f_{\bar N} \f_{\bar N}(x) \ot \hat T_H(x)
\label{eq:shs}
\ee
and this is related to the partial decay width by $\Gamma \sim |\cm|^2 /M$.
So $\cm$ must have mass dimension one. On the right hand side of 
\eref{eq:shs}, the only dimensional quantities are the decay 
constants $f$'s and a certain power of $1/M$ hidden in $\hat T_H$.
They must make up the right dimension to match the left hand side.
The decay constants have the following mass dimension.
$f_N$, $f_{\bar N}$ and $f^8_{\c_J}$ the octet constant are from 
3-particle wavefunctions and so must have mass dimension two. 
The singlet constant $f^0_{\c_J}$ is from a 2-particle wavefunction 
which would mean dimension one but being a P-wave increased this to two. 
Now extracting sufficient power of $1/M$ from $\hat T_H$ to make up
the right dimension for $\cm$ gives
\bea \cm^0 \fx & \sim & \fx 
     M \frac{f^0_{\c_J}}{M^2} \Big ( \frac{f_N}{M^2} \Big )^2  
     \sim \frac{1}{M^5}   \\
     \cm^8 \fx & \sim & \fx 
     M \frac{f^8_{\c_J}}{M^2} \Big ( \frac{f_N}{M^2} \Big )^2  
     \sim \frac{1}{M^5}   \; .
\eea
We see that both the colour singlet and octet are weighed by the 
inverse fifth power of the large scale $M$. Therefore the colour 
octet is not suppressed when compared to the singlet contribution. 

Now we are ready to return to the $1/M$ suppression of P-wave wavefunction
discussed in the previous section. Only because of this suppression is
the singlet amplitude scale like $1/M^5$. If this were not present
as in the case of the S-wave $J/\j$, the singlet amplitude would scale
like $1/M^4$ instead. In that case, it would be legitimate
to neglect the colour octet contribution.

\section{Calculation --- singlet contribution}

To obtain the colour singlet contribution to the decay width,
instead of the original hard scattering scheme, we employ the 
modified version of Botts, Li and Sterman \cite{bs,ls}.
In this modified scheme, the probability amplitude is now
a convolution of wavefunctions, hard perturbative part $\hat T_H$
and the exponential Sudakov suppression factor 
\bea \cm \fx &\sim &\fx  \j_{\c_J}(x,\kt) \ot \j_N(x,\kt) 
     \ot \j_{\bar N}(x,\kt)                   \nonumber \\               
     \fx & & \fx \ot \hat T_H(x,\kt,\a_s(x)) \ot \exp \{-S(x,\kt) \}   
     \; .
\label{eq:mhs}
\eea
The essential differences between this and the standard scheme are
that internal transverse momenta are kept everywhere, the strong
coupling $\a_s$ is part of the convolution instead of a constant
and Sudakov suppression is included. Actually even in the
standard scheme, $\a_s$ need not be a constant and can be made 
part of the convolution, but in that case, some adhoc arrangement 
such as freezing the value of the coupling at small momentum 
transfer will have to be made. The modified scheme has no such
problem as the more natural Sudakov factor takes care of 
potential divergence of the coupling. This is achieved by
using the adjacent largest virtuality in the neighbourhood
of each coupling for its argument \cite{bks1,bks2}. It can be 
that of the neighbouring gluon propagator, quark propagator or the 
inverse internal transverse separation of quark pairs in the 
outgoing nucleon-antinucleon. In this way, the scale of the 
decay process is determined dynamically. 

Our objection to some of the calculations based on using 
the standard scheme is that if a fixed coupling is not
used then freezing of the coupling or one of the other methods to 
deal with this is necessary. These methods have not a sound physical
basis. If a constant coupling is used, then taking our current
example of $\c_J$ decay into nucleon-antinucleon the probability
amplitude goes like $\cm \sim  \a_s^3$ and the decay width
is therefore $\Gamma \sim \a_s^6$. A choice of $\a_s = 0.3$
instead of $\a_s =0.5$ will therefore make a factor of 20 
difference in the width so provided the singlet countribution is
of reasonable size, one can always argue for a value of the
coupling so that the singlet contribution is sufficient
to account for the measured partial width. This in our
opinion is quite arbitrary and unsatisfactory

Using the modified scheme, we found that the colour singlet 
component makes up of no more than 6 and 12 \% of the measured 
decay widths for $\c_1$ and $\c_2$ respectively.
So colour singlet alone is not sufficent in agreement 
with our theoretical expectation.

\section{Calculation --- including octet contribution}

To complete the calculation, one must include the colour octet
contribution. It turned out that to use the modified scheme
to include also the octet contribution was rather daunting
because the internal transverse momentum must be kept everywhere
and with an extra constituent gluon, the dimension of the
numerical integrations became quite high. One could go back
to the standard scheme but then had to face the objections
raised in the previous section. Since the problem with
using the standard scheme is mainly in the coupling, one
can circumvent this by using a model coupling that is free of
the Landau pole and thus can be included in the convolution
of probability distribution amplitudes. With such a model,
the $\c_J$ decay probability amplitude into $N\bar N$ becomes
\bea \cm \fx & \sim & \fx f_{\c_J} \f_{\c_J}(x) \ot f_N \f_N(x) 
     \ot f_{\bar N} \f_{\bar N}(x)                     \nonumber \\
     \fx & & \fx 
     \ot \hat T_H(x,\a^{\mbox{\fiverm model}}_s(x))   \; .
\label{eq:smhs}
\eea
There exist various models that fit our requirement. 
We choose the one constructed by Shirkov 
and Solovtsov \cite{ss} which is less complicated and the Landau 
pole is removed by simple subtraction. To one-loop, it is 
\be \a^{\mbox{\fiverm model}}_s(\m) = \frac{4\p}{\b_0} 
    \Big \{\frac{1}{\ln \m^2/\Lambda^2}+\frac{\Lambda^2}{\Lambda^2-\m^2}
    \Big \}  \; .
\ee
Although this model is comparatively simple, it agrees with
known estimates of the quantity 
\be A(Q) = \frac{1}{Q} \int^Q_0 \dd k \; \a_s(k)
\ee
from jet physics. These estimates are done at a value of $Q\sim 2$ GeV
\cite{dw,dkt}. 

\begin{table}
\caption{Prelimary estimates of the colour singlet plus octet
contributions to $\c_J$ decay into $N\bar N$ in our scheme.}
\label{t:est}
\begin{center}
\begin{tabular}{ccc}
\hline
  $J \; \; \; \;$ & $\Gamma(\c_J\rightarrow N\bar N)$ [eV] \ \ \ \
      & PDG [eV] \\ \hline 
 1 \ \  & \ \ 89.0  & \  75.68   \\ 
 2 \ \  &    180.0  &   200.00   \\ \hline
\end{tabular}
\end{center}
\end{table}

Without the $\kt$ dependence, the calculation becomes much simpler
and with our so constructed semi-modified hard scattering scheme,
the dynamical setting of scale by the process itself is preserved.
Details of our calculation parallel those of \cite{bks1,bks2}.
Our preliminary results of the total colour singlet plus octet
contribution together with the experimental measurements 
\cite{pdg} are shown in Table 1 \cite{kw}. As can be seen, the 
agreement with the measured values are quite reasonable.
This confirm our theoretical argument in Sec. \ref{s:svso}.
In fact, this argument can be extended to even higher wave
quarkonia and in such cases, not only the next higher Fock
state must be included but also the next-next higher states. 

In view of the above considerations and results, the constituent 
gluon in P-wave quarkonia is an important
part of the heavy quark hadrons. The description that they are 
bounded heavy quark-antiquark systems becomes inaccurate. If the 
\vfill
\eject
\noindent ``quark'' in quarkonium means
heavy quark-antiquark object, then the three $\c_J$'s would each
be less of a quarkonium than a $J/\j$ and a D-wave would be even 
less so. Therefore the nomenclature is missleading in this sense.

\end{document}